# Femtosecond photoexcited carrier dynamics in reduced graphene oxide suspensions and films

**Sunil Kumar,[1,2] N. Kamaraju,[1,2] K. S. Vasu,[1] A. K. Sood[1,2,*]**

[1]*Department of Physics, Indian Institute of Science*

*Bangalore 560 012, India*

[2]*Center for Ultrafast Laser Applications*

*Indian Institute of Science, Bangalore 560 012, India*

**asood@physics.iisc.ernet.in*

We report ultrafast response of femtosecond photoexcited carriers in single layer reduced graphene oxide flakes suspended in water as well as few layer thick film deposited on indium tin oxide coated glass plate using pump-probe differential transmission spectroscopy at 790 nm. The carrier relaxation dynamics has three components: ~200 fs, 1 to 2 ps, and ~ 25 ps, all of them independent of pump fluence. It is seen that the second component (1 to 2 ps) assigned to the lifetime of hot optical phonons is larger for graphene in suspensions whereas other two time constants are the same for both the suspension and the film. The value of third order nonlinear susceptibility estimated from the pump-probe experiments is compared with that obtained from the open aperture Z-scan results for the suspension.

*Keywords*: Graphene; femtosecond pump-probe spectroscopy; nonlinear properties.

## 1. Introduction

The gapless energy spectrum of electrons and holes with linear dispersion relation up to a few eV at K points (Dirac points) in graphene at equilibrium leads to unique features in electrical transport and optical properties. Dynamics associated with carrier intraband relaxation and interband recombination have been studied using ultrafast pump-probe spectroscopy of ultra-thin film of graphite [1], epitaxial graphene [2-5] and graphene suspensions [6]. A faster component of ~ 100 fs has been attributed to intraband carrier-carrier scattering and a slower component between 0.4 ps to 5 ps to intraband carrier-phonon scattering [2,6]. The latter depends on the defect density in the sample - being smaller for more defected sample [2,4]. In other studies on ultrathin graphitic film of thickness 17 nm [1] and epitaxial graphene grown on SiC [5], it was shown that photoexcited carriers loose most of their energy within first 500 fs by emitting optical phonons resulting in creation of hot optical phonons which survive up to a few ps and present the main bottleneck to the subsequent cooling of the carriers.

Here, we report the results on relaxation dynamics of photoexcited carriers in reduced graphene oxide (RGO) flakes suspended in water (RGO-suspension) and deposited on indium-tin-oxide coated glass plates (RGO-film). The suspension contains mostly single layer graphene as evident from a single '2D' band at 2686 $cm^{-1}$ in the Raman spectra [7]. There are three time constants associated with the decay of the photoexcited carriers: first one ~ 200 fs, second one between 1 to 2 ps and third one ~ 25 ps, all of them nearly independent of the pump fluence. The first and third components are attributed to electron-optical phonon and electron-acoustic phonon scattering, which are found to be similar for the suspension and the film. The second time constant, on the other hand, which is attributed to the lifetime of optical phonons in graphene, is smaller for the RGO-film. We estimate the nonlinear absorption coefficient from the degenerate pump-probe signal and compare it with that obtained from open aperture Z-scan measurements for the RGO-suspension.

## 2. Experimental Procedure

RGO-suspension having graphene concentration of ~ 80µg/ml was prepared and characterized as reported in Ref. [7]. 150µl of the suspension taken in a 1 mm optical path length quartz cuvette was used in the degenerate pump-probe and Z-scan experiments. To make film of RGO (thickness ~ 75 nm), 20 µl of the suspension was dropped on indium tin oxide (ITO) coated glass plate and allowed to dry for a day. Degenerate pump-probe experiments at 790 nm were performed on the suspension and the film using 80 fs laser pulses from a Ti:sapphire laser amplifier (1kHz, Spitfire, Spectra Physics). The pump fluence was varied between 0.12mJ/cm$^2$ and 1.21mJ/cm$^2$ whereas the probe fluence was kept constant at 9.6µJ/cm$^2$. The pump beam was modulated at a frequency of 383Hz using an optical chopper and the change in probe transmitted intensity due to the presence of the pump was recorded using a Si-PIN diode in a lock-in amplifier detection scheme.

## 3. Results and Discussion

Results from degenerate pump-probe experiments for the RGO-suspension and RGO-film are presented in Fig. 1 where $T$ is the linear transmission of the probe beam in absence of the pump and $\Delta T(t)$ is the change in probe transmission in presence of the pump as a function of time delay ($t$) between the pump and probe pulses. Fig. 1 shows that the magnitude of the signal at zero delay is higher for the RGO-suspension as compared to RGO film at the same pump fluence. The magnitude of positive $\Delta T$ signal at zero delay corresponds to hot thermal carrier distribution which is established immediately after photo-excitation within a time scale comparable to the laser pulse width. This nonthermal carrier distribution relaxes towards equilibrium distribution on a time scale of ~ 25 ps for both the suspension and the film. It can be seen from Fig. 1 that the signal decreases rapidly to almost 15% of its peak value in the first 250 fs and then decreases much slowly afterwards.

The differential transmission signal is fitted with the following function:

$$\frac{\Delta T}{T} = \left[ H(t) \sum_{i=1}^{3} A_i \exp(-t/\tau_i) \right] \otimes G(t) \qquad (1)$$

Here $H(t)$ is the Heaviside step function and resultant is convoluted with Gaussian cross-correlation function $G(t)$ to account for the laser pulse width. The fitted results are shown by solid lines in Fig. 1.

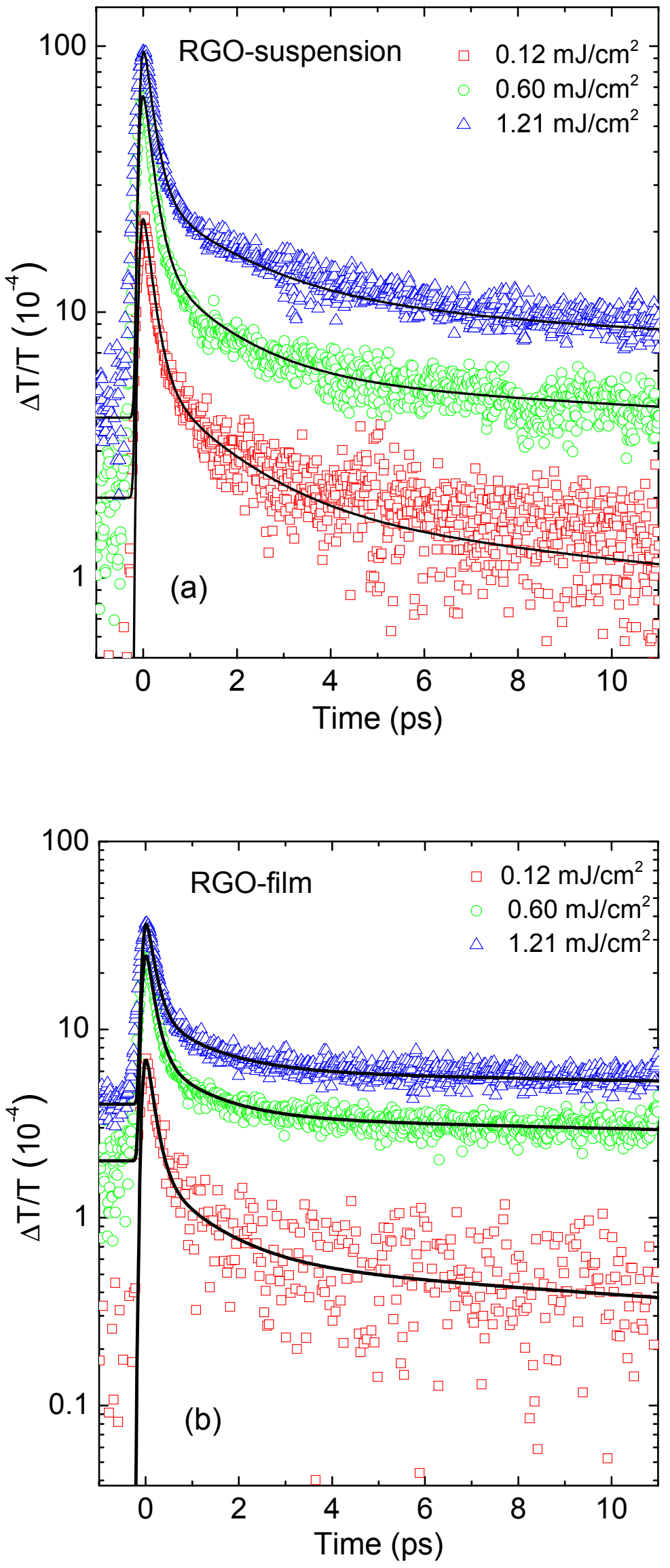

Fig. 1. Differential transmission spectra for (a) RGO-suspension, and (b) RGO-film using degenerate pump-probe measurements at 790 nm. The pump fluence was varied from 0.12mJ/cm$^2$ to 1.21mJ/cm$^2$ whereas the probe fluence was constant at 9.6µJ/cm$^2$.

The magnitudes and associated time constants of the three components as obtained from the fits are presented in Fig. 2 for the suspension (open circles) and the film (filled circles). It can be seen that the coefficients, $A_i$ and time constants, $\tau_i$ are nearly independent of pump fluence. $\tau_1$ and $\tau_3$ are similar for both the suspension and the film whereas $\tau_2$ is larger for the suspension than the film. The time constant $\tau_2$ is attributed to the lifetime of the optical phonons in graphene as discussed in the next paragraph. In the suspension, the graphene layers are isolated from each other and hence the generated optical phonons survive for a longer time. On the other hand, in case of the film, layer to layer interaction and interaction with the substrate open up new channels for the phonons to release their energy to low energy acoustic phonons.

The number density of photoexcited carriers is a function of pump fluence and determines the interband electron-hole recombination times [4]. We note from Fig. 2 that $\tau_3$ is nearly independent of the pump fluence and hence can not be attributed to carrier recombination. Since the defect levels need not be same for the RGO in suspension and the film, $\tau_3$ being almost the same for both of them can be associated with the carrier-acoustic phonon scattering in graphene rather than carrier-defect scattering.

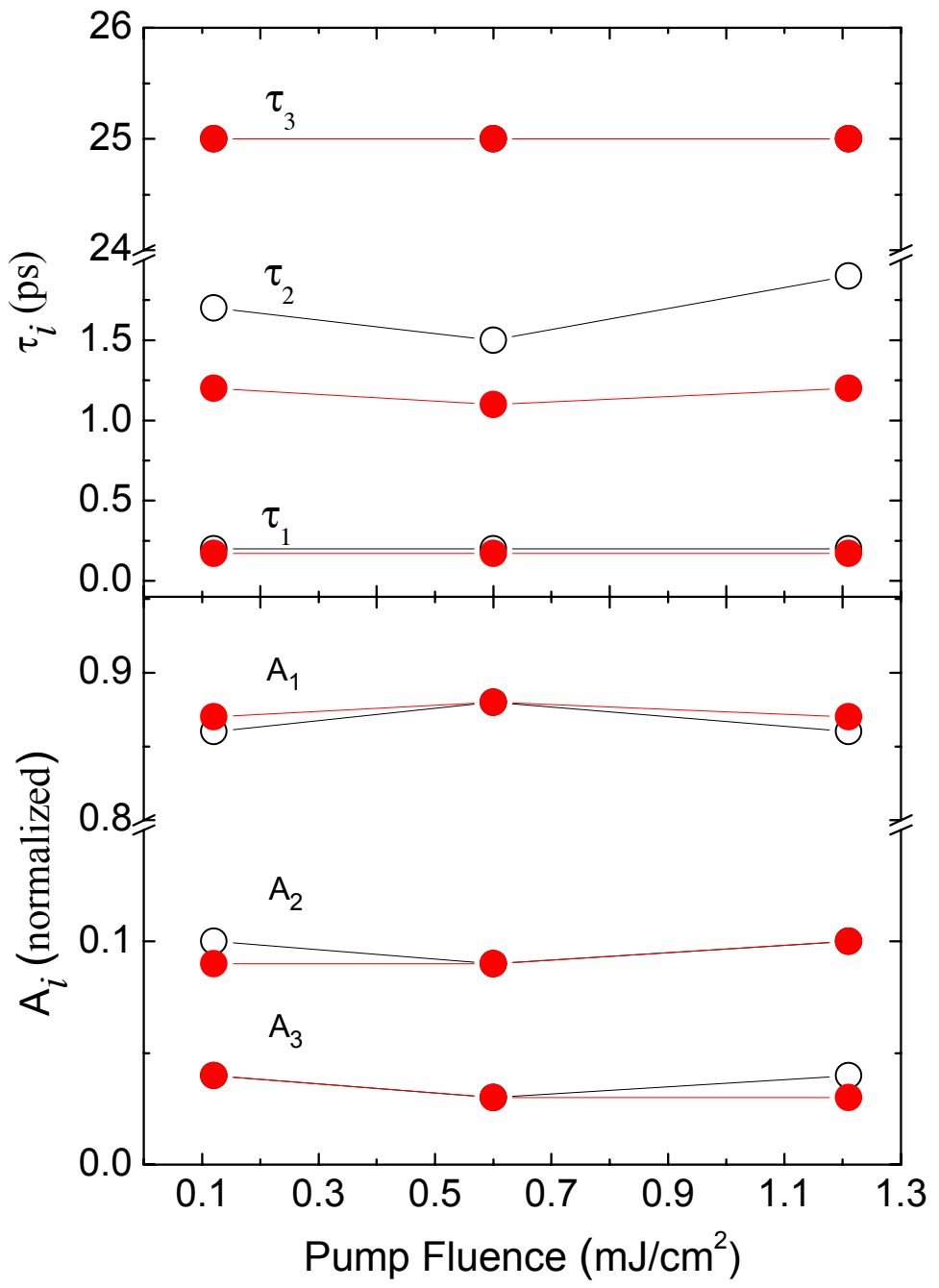

Fig. 2. Comparison between the amplitude of three components and decay time constants of photoexcited carriers in RGO-suspension (open circles) and RGO-film (filled circles).

The photoexcited carrier density can be approximately calculated as $n \sim F\alpha d/h\nu$, where $F$ is the pump fluence, $\alpha$ is the linear absorption coefficient and $h\nu$ is the pump photon energy. For RGO-suspension, $\alpha \sim 1.0\text{x}10^{4}$ cm$^{-1}$, and considering, the thickness of the graphene sheet, $d \sim 1$ nm, we obtain $n \sim 5\text{x}10^{12}$ cm$^{-2}$ at $F = 1.2$ mJ/cm$^2$ which corresponds to the initial carrier temperature of ~ 3030 K. At such level of carrier densities, the carrier mean free path, $l$ has been theoretically estimated to be ~ 50 nm, which in turn corresponds to carrier scattering time $\tau \sim l/v_F$ (Fermi velocity $v_F \sim 10^8$ cm/s) of the order of 50 fs [8]. For samples having defects, this time constant is expected to be even shorter. In our present experiments, the fastest component, $\tau_1 \sim 200$ fs, is much larger than 50 fs for both the suspension and the film. This leads us to associate $\tau_1$ with electron-optical phonon scattering instead of electron-electron scattering as suggested in earlier studies on epitaxial graphene [3,5] and ultrathin film of graphite [1].

Optical excitation by ~1.6 eV (790 nm) laser pulse creates a nonthermal carrier population 800 meV above the band extrema. The initial relaxation takes place within $\tau_1 \sim 200$ fs and slows down later, resulting in a bottleneck which can be explained due to two possible reasons [9]; firstly, for the relaxation from conduction band to valence band all the electrons have to cross the vicinity of the K point which has a low density of states and therefore acts as a bottleneck for the dynamics. Secondly, the early relaxation leads to emission of cascade of G-band (~193 meV) optical phonons [1,3,9] which survive for $\tau_2 \sim 1.8$ ps in the suspension and ~ 1.2 ps in the film. These numbers for $\tau_2$ are close to the life time of G-phonons in graphite (~ 2.2 ps) measured using time-resolved Raman spectroscopy [10]. Since the efficient channels are now occupied after heating the optical phonons which further split into acoustic phonons, a large time scale for electron-acoustic phonon scattering, $\tau_3 \sim 25$ ps is observed for both the suspension and the film.

The positive sign of $S = \Delta T/T$ at $t = 0$ is generally referred to as bleaching of the ground state of the system or saturable absorption in the degenerate pump-probe experiments. Fig. 3 compares the values of $S$ for both the suspension and the film at three pump fluences. It is seen that $S$ for the film is less than half as compared to the suspension. The magnitude of the nonlinear absorption coefficient, $\beta$ can be approximately determined from $S$ by using the relation [6] $\beta = \ln(1+S)/(L_{\text{eff}} I_{\text{pump}})$, where $L_{\text{eff}}$ = (cuvette optical length x graphene concentration) is the effective optical path length of the sample (~ 80 nm for the suspension and the film) and $I_{\text{pump}}$ is the pump intensity. The parameter $\beta$, in turn, is related to the imaginary part of third order nonlinear optical susceptibility, $\text{Im}\chi^{(3)} = [10^{-7} c\lambda\eta^2/96\pi^2]\beta$, where $\text{Im}\chi^{(3)}$ is in esu, $\beta$ is in cm/W, $c$ is speed of light in vacuum, $\lambda$ is the laser wavelength and $\eta$ is the linear refractive index of the sample (taken as 1.5). At the maximum pump fluence of 1.21 mJ/cm$^2$, the calculated values of $\beta$ and $\text{Im}\chi^{(3)}$ for RGO-suspension are $7.5\text{x}10^{-8}$ cm/W and $4.2\text{x}10^{-11}$ esu. For the RGO-film, these values are $2.9\text{x}10^{-8}$ cm/W and $1.6x10^{-11}$ esu, almost 2.5 times smaller than that for the suspension.

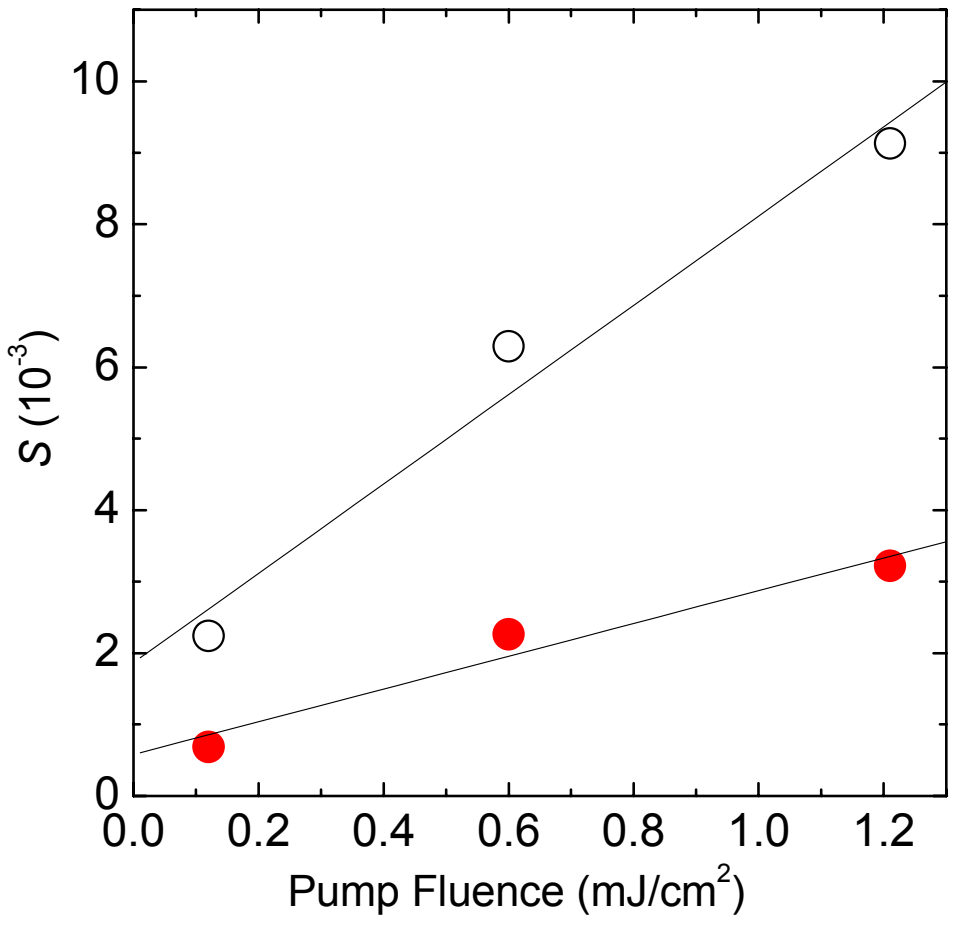

Fig. 3. Comparison between the maximum of Differential transmission spectra for RGO-suspension (open circles) and RGO-film (filled circles). The solid lines are linear fits.

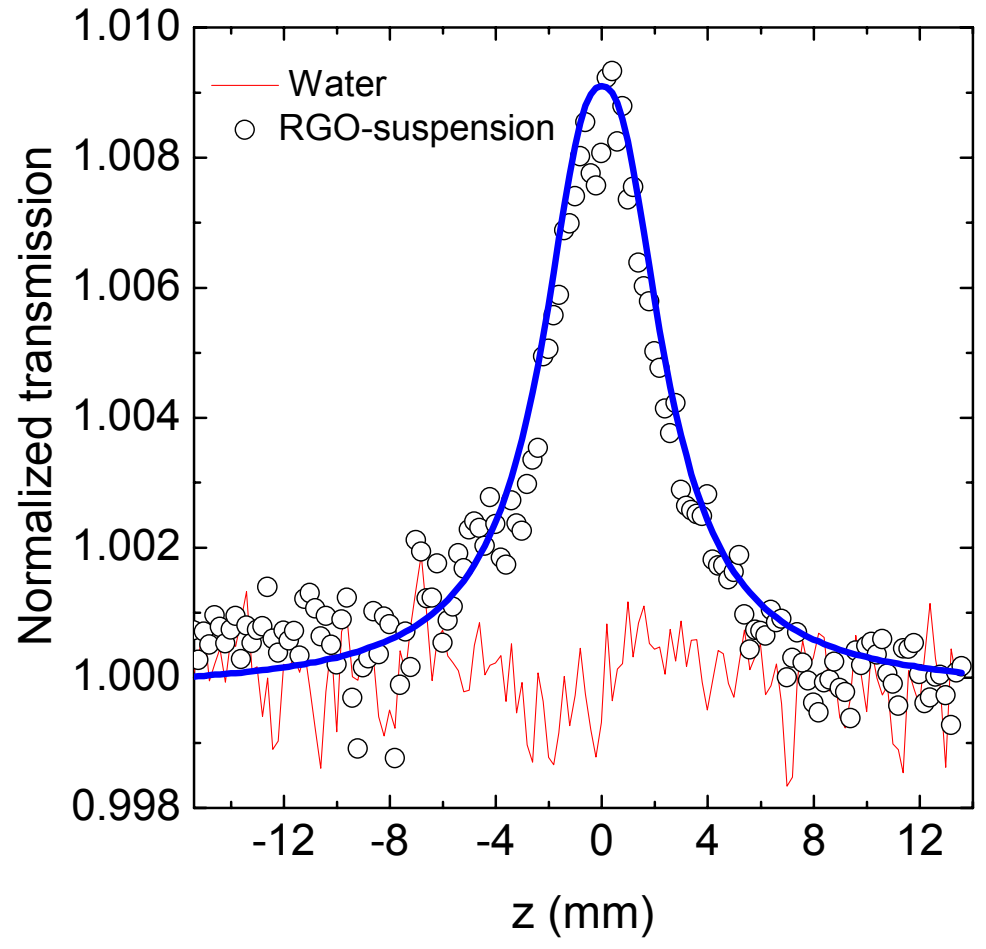

Fig. 4. OA Z-scan data and fit for the RGO-suspension.

In Fig. 4 we show the results from open aperture (OA) Z-scan measurements on the RGO-suspension and water only, taken in a 1 mm optical path length cuvette. A saturable absorption behavior for the RGO can be clearly seen. The data has been fitted (solid line) with the model given in Ref. [11] to deduce nonlinear optical constant, $\beta$ and the saturation intensity $I_s$ as $\beta = 4.5x10^{-8}$ cm/W and $I_s$ = 100 GW/cm$^2$. A figure of merit (FOM), defined as FOM = $|\text{Im}\chi^{(3)}|/\alpha$ gives FOM = $4.5x10^{-15}$ esu.cm.

## 4. Conclusion

In summary, we have studied the ultrafast response after femtosecond pulse excitation of single layer reduced graphene oxide flakes suspended in water as well as few layer thick film deposited on ITO coated glass plate. Three component relaxation dynamics shows a fast component ~ 200 fs, second component between 1 and 2 ps and a slower one ~ 25 ps. The assignment of these components is discussed in our work.

## Acknowledgments

A.K.S. thanks Department of Science and Technology for financial support and S.K. acknowledges University Grants Commission for Senior Research Fellowship.